\documentclass[12pt,preprint]{aastex}

\slugcomment{To appear in AJ.}

\shorttitle{Herbig-Haro Objects in \object{NGC 7023} and \object{B175}}
\shortauthors{Rector \& Schweiker}

\begin{document}


\title{A Search for Herbig-Haro Objects in \object{NGC 7023} and Barnard~175}


\author{T.A. Rector}
\affil{Department of Physics and Astronomy, University of Alaska Anchorage,
    Anchorage, AK 99508}
\email{rector@uaa.alaska.edu}

\and

\author{H. Schweiker\altaffilmark{1,2}}
\affil{National Optical Astronomy Observatory, Tucson, AZ 85719}
\email{heidis@noao.edu}


%\altaffiltext{}{rector@uaa.alaska.edu}
\altaffiltext{1}{Kitt Peak National Observatory, National Optical Astronomy Observatory, which is operated by the Association of Universities for Research in Astronomy, Inc. (AURA) under cooperative agreement with the National Science Foundation.}
\altaffiltext{2}{WIYN Observatory, Tucson, AZ 85719}
%\altaffiltext{4}{heidis@noao.edu}


\begin{abstract}
Wide-field optical imaging was obtained of the cluster and reflection nebula \object{NGC 7023} and the Bok globule \object{B175}.  We report the discovery of four new Herbig-Haro (HH) objects in NGC~7023, the first HH objects to be found in this region.  They were first detected by their H$\alpha$ and [S~II] emission but are also visible at 3.6 and 4.5\micron\ in archival {\it Spitzer} observations of this field.  These HH objects are part of at least two distinct outflows.  Both outflows are aligned with embedded ``Class I" YSOs in a tight group on the western edge of the nebula.  One of the outflows may have a projected distance of 0.75pc, which is a notable length for an embedded source.

No new HH objects were discovered in B175.  However, we reclassify the knot \object{HH450X}, in B175, as a background galaxy.   The discovery that HH~450X is not a shock front weakens the argument that HH~450 and SNR~G110.3+11.3 are co-located and interacting.
\end{abstract}


\keywords{ISM:  jets and outflows --- Herbig-Haro objects --- ISM: individual objects (NGC 7023, B175) --- stars: formation}



\section{Introduction}

Supersonic outflows from young stellar objects (YSOs) are often seen at optical wavelengths as Herbig-Haro (HH) objects.  HH objects are produced by shocks from the collision of the outflow with the external medium or with previously ejected outflow components, producing collisionally excited forbidden lines.  In the optical, they can be readily identified by their H$\alpha$ and [S~II] line emission, non-stellar appearance, and nearly collinear alignment.  See, e.g., \citet{bal01a} for a more detailed discussion.  Despite their distinctive characteristics, HH objects can be difficult to find in the dense, complex environment in which YSOs are forming.  In this paper we report on the search for HH objects in the regions of \object{NGC 7023} and \object[B175]{Barnard 175} (B175).

\object{NGC 7023} is a reflection nebula illuminated by the Herbig Be star \object{HD 200775}.  It is embedded in a giant molecular cloud complex known as the ``Cepheus Flare" Shell \citep{hub34} and is on the northern edge of the $L1167/L1174$ complex.  The measured distance to HD~200775 via {\it Hipparcos} parallax is $430^{+160}_{-90}$ pc \citep{van97}, which we adopt here; although \citet{kun08} and references therein argue for other distances.  No prior discoveries of HH objects in NGC~7023 have been reported.

B175, which is also embedded in the Cepheus Flare, is a cometary-shaped Bok globule. It is illuminated on the southern edge by the B9.5V type star \object{BD+69 1231}, producing the reflection nebula \object{Ced 201}.  The distance to B175 is estimated to be at a distance of about 400pc; see \citet{bal01b} for discussion.  \citet{bal01b} also report the discovery of \object{HH~450} embedded in Ced~201.  It is the only known HH object within B175.  

\section{Observations}

%% In a manner similar to \objectname authors can provide links to dataset
%% hosted at participating data centers via the \dataset{} command.  The
%% second curly bracket argument is printed in the text while the first
%% parentheses argument serves as the valid data set identifier.  Large
%% lists of data set are best provided in a table (see Table 3 for an example).
%% Valid data set identifiers should be obtained from the data center that
%% is currently hosting the data.
%%
%% Note that AASTeX interprets everything between the curly braces in the 
%% macro as regular text, so any special characters, e.g. "#" or "_," must be 
%% preceded by a backslash. Otherwise, you will get a LaTeX error when you 
%% compile your manuscript.  Special characters do not 
%% need to be escaped in the optional, square-bracket argument.

NGC~7023 and B175 were observed with the MOSAIC camera on the Mayall 4-meter telescope at Kitt Peak National Observatory.  MOSAIC is an optical camera that consists of eight 2048$\times$4096 CCD detectors arranged to form a 8192$\times$8192 array with 35 to 50-pixel-wide gaps between the CCDs.  With a scale of $0\farcs26$ pixel$^{-1}$, the field of view is approximately $36\arcmin\times 36\arcmin$.  To fill in the gaps and bad columns, all observations are completed in a five-exposure dither pattern with offsets of approximately 100 pixels.  

Observations of NGC~7023 were obtained on 29 August 2009 with the Harris $B$ (MOSAIC filter k1002), Harris $V$ (k1003), ``Nearly-Mould" $I$ (k1005) and H$\alpha$ (k1009) filters.  Five exposures each of 600~sec in $B$, 420~sec in $V$, 300~sec in $I$ and 600~sec in H$\alpha$ were obtained.  Excellent seeing of $0\farcs85$ in H$\alpha$ was achieved.  After the outflows were first detected in NGC~7023, follow-up observations were completed in [S~II] (ha16, H-alpha+16nm, k1013) on 21 June 2012.  Five exposures of 600~sec in [S~II] were obtained.  Unfortunately the seeing was $1\farcs8$; but it was sufficient for confirmation purposes.

Observations of B175 were obtained on 23 October 2011 with the  $B$ (k1002), $V$ (k1003), $I$ (k1005) and H$\alpha$ (k1009) filters.  Five exposures each of 480~sec in $B$, 300~sec in $V$, 180~sec in $I$ and 600~sec in H$\alpha$ were obtained.  Seeing of 1\farcs1 in H$\alpha$ was achieved.  

At the time of the August 2009 observations MOSAIC was equipped with thinned, science-grade SITe CCD cameras.  In the summer of 2012  MOSAIC was upgraded to eight thinned, science-grade e2v CCDs.   The observations of B175 and follow-up [S~II] observations of NGC~7023 were therefore obtained after the upgrade.  Both CCD types have 15\micron\ pixels; thus the pixel scales of the CCDs are nearly identical.  And the DQEs of the two cameras are very similar at the wavelength range of interest (at 6500\AA, an average of 86\% for the SITe chips and 80\% for the e2v chips).  The bandpasses of the H$\alpha$ (K1009) and [S~II] (k1013) filters are $\sim80$\AA, with transmission efficiencies of $\sim90\%$ at $\lambda 6563$ and $\lambda\lambda 6717,6731$ respectively.  Thus they should be roughly equivalent in sensitivity to H$\alpha$ and [S~II] emission.

The data were reduced with the IRAF package MSCRED in the standard manner.  Ten bias frames and five dome flats in each filter were used.  The world coordinate system (WCS) was determined via stars from the USNO-B1.0 catalog \citep{mon03} with a global solution RMS of better than $0\farcs4$ in all cases. This is assumed as the accuracy for all measured positions.  All of the images were projected to a common WCS to remove geometric distortions and to allow for the images to be aligned.  The images in each filter were then combined to fill in the gaps.

%% In this section, we use  the \subsection command to set off
%% a subsection.  \footnote is used to insert a footnote to the text.


%% This section also includes several of the displayed math environments
%% mentioned in the Author Guide.

\section{Results}

As a means of visualizing the structure, the data for each object were combined to form a color-composite image with the methodology described in \citet{rec07}.  An advantage of searching for HH objects in a color-composite image is that the H$\alpha$ emission is distinct and readily visible if assigned a unique color.  Further, the broadband filters reveal the amount of obscuration from dust and gas.  Thus, faint outflows can be found more easily in the complex environments typical of star-forming regions. The results for each object are discussed below.

\subsection{NGC~7023}

\begin{figure}
\plottwo{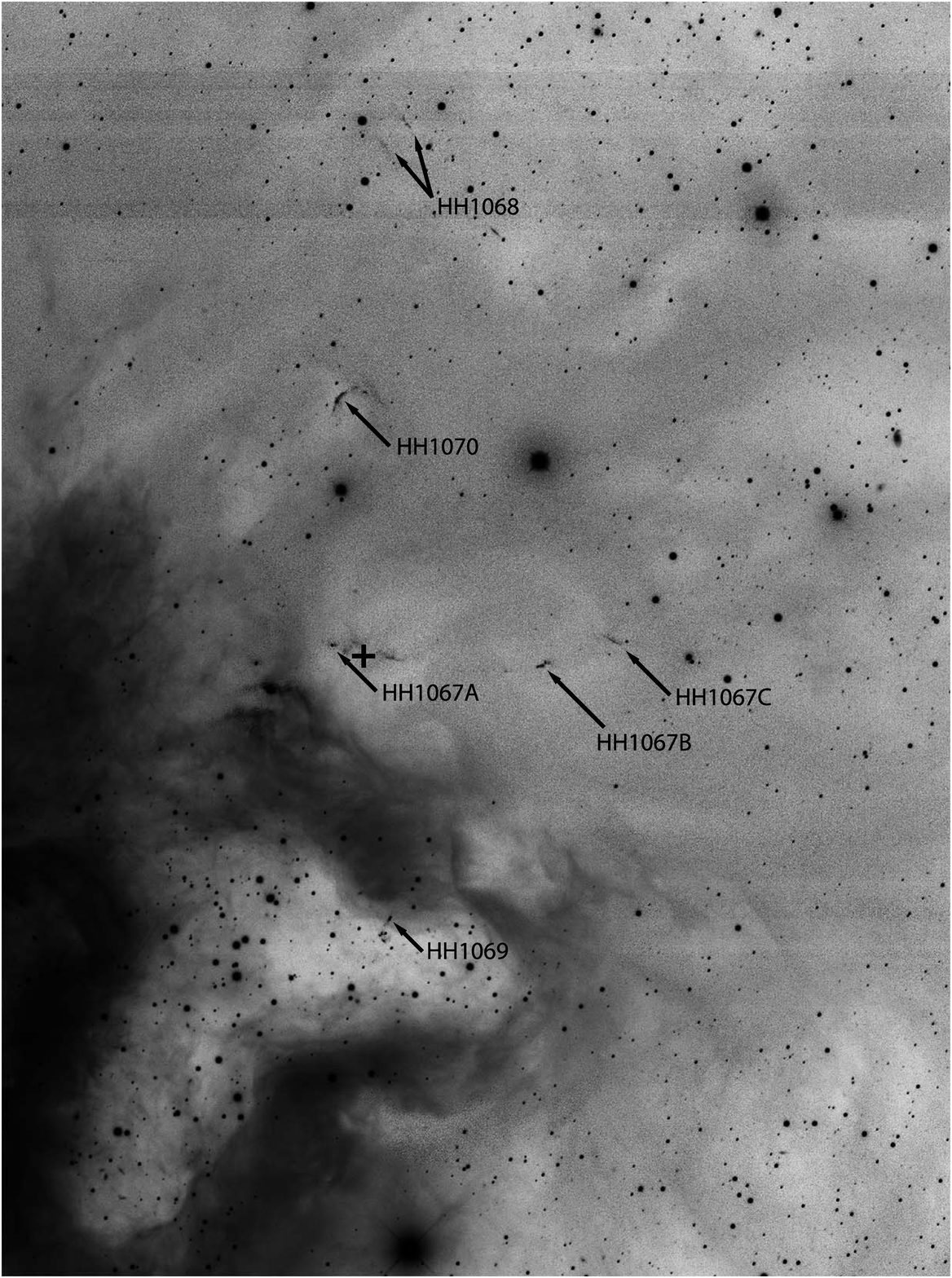}{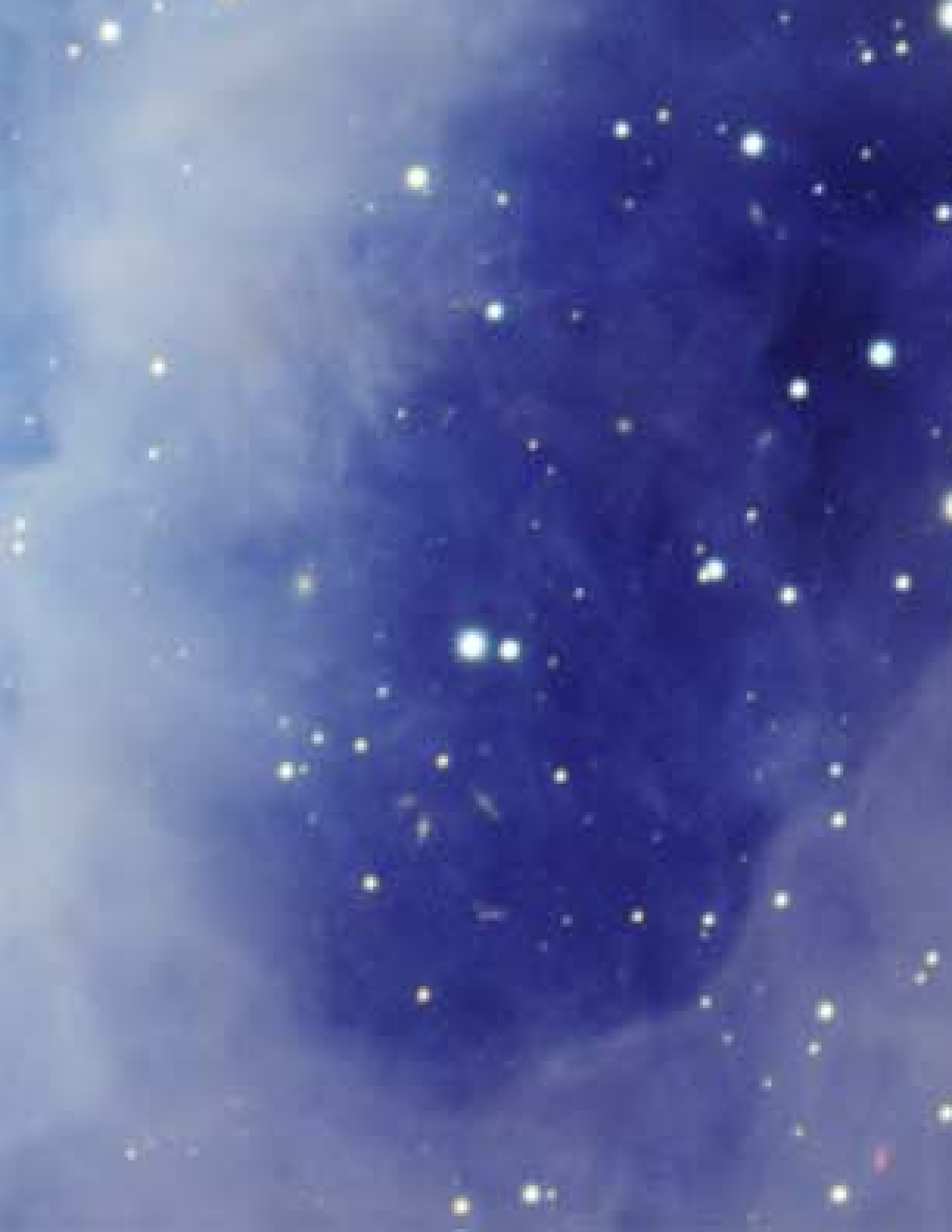}
\caption{An optical color composite image of the western portion of NGC~7023 that contains the four newly discovered HH objects.  The field of view is 10\farcm7$\times$14\farcm4.  North is up, east is to the left.  
The HH objects are labeled, as is the location of Tight Group B (shown as a cross).   In the grayscale image, the color red was enhanced to better show the HH objects.  See the electronic 
edition of the Journal for a color version of this figure.  In the color version, the color assignments for the filters are: $B$ (blue), $V$ (green), $R$ (orange) and H$\alpha$ (red).
\label{fig-1}}
\end{figure}

\begin{figure}
\plotone{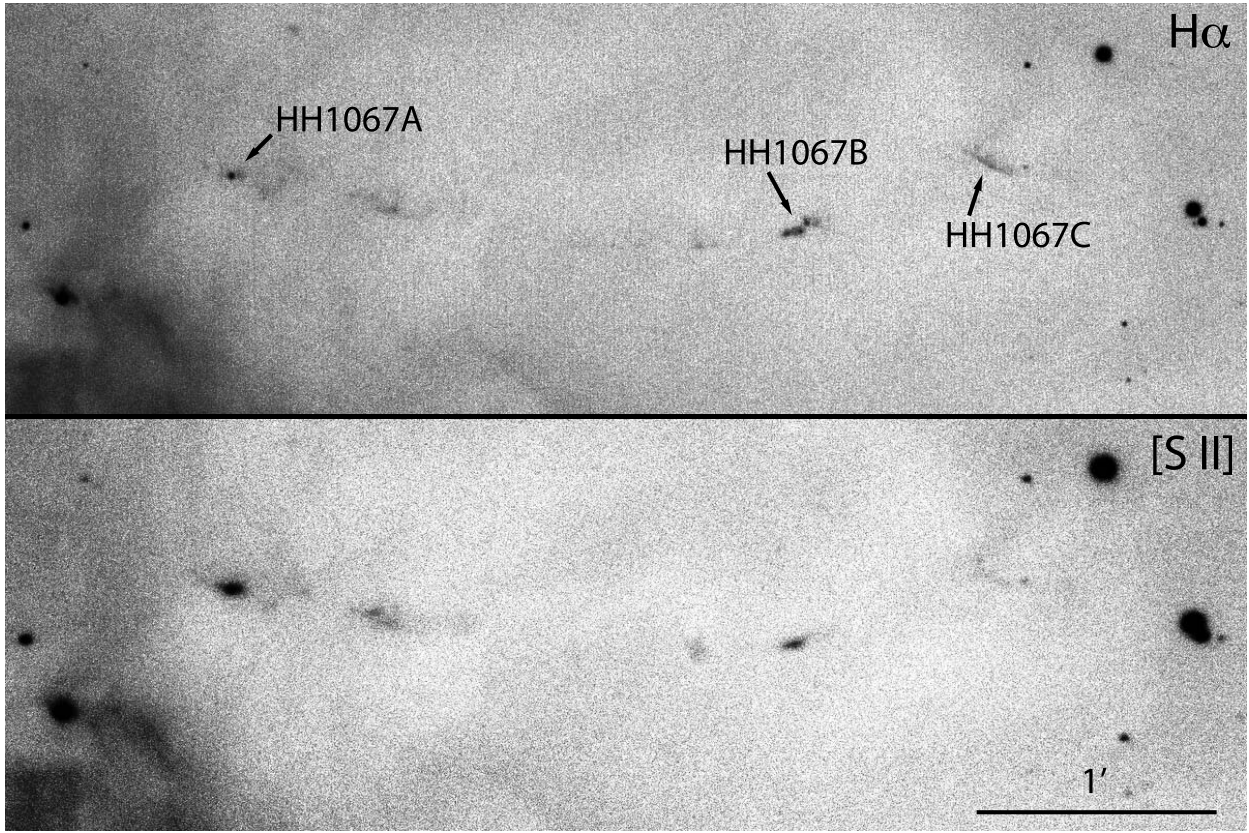}
\caption{HH1067 in H$\alpha$ (top) and [S~II] (bottom).  The filament that connects HH1067A and HH1067B is visible in both filters.  Note that over three years have elapsed between the two images; and that the seeing is poor in the [S~II] image.\label{fig-2}}
\end{figure}

\begin{figure}
\plotone{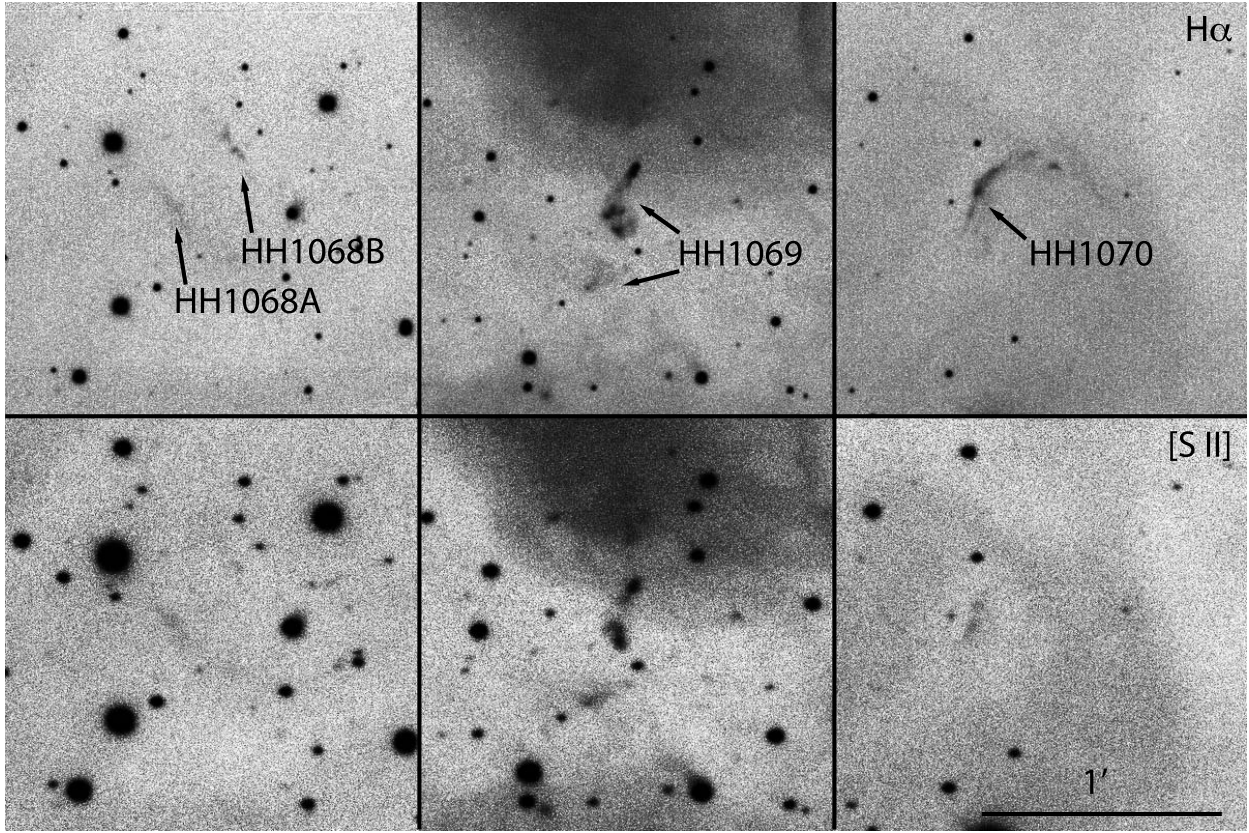}
\caption{HH1068, HH1069 and HH1070 in H$\alpha$ (top) and [S~II] (bottom).  HH1068 is only faintly detected in both filters.  HH1069 is bright in both filters.  And only the eastern edge of HH1070 is detected in [S~II].  Note that over three years have elapsed between the two images; and that the seeing is poor in the [S~II] image.\label{fig-3}}
\end{figure}

A region 36\farcm5 $\times$ 36\farcm3 in size, centered on the location ${\rm \alpha}_{2000} =21^{\rm h}01^{\rm m}30\fs5, {\rm \delta}_{2000} = +68\degr09\arcmin57\arcsec$ was searched for new HH objects.  At least two distinct outflows were discovered in the northwestern ``lobe" of NGC~7023.  Four new HH objects are labeled in Figure~\ref{fig-1}.  All of the HH objects are visible in the H$\alpha$ and [S~II] filters, but are not detected in the broadband filters.  Thus we are confident they are sources of line emission only.  The objects were named in order of increasing right ascension.  When objects are clearly part of the same outflow (e.g., HH~1067) they share a single HH number.  When it is uncertain if individual knots are part of the same flow, they are given separate HH numbers.  Their coordinates are given in Table~\ref{tbl-1}.  The given positions correspond to the brightest knot in each object.  Figures~\ref{fig-2} and \ref{fig-3} show the H$\alpha$ and [S~II] images for each object.

\begin{figure}
\plottwo{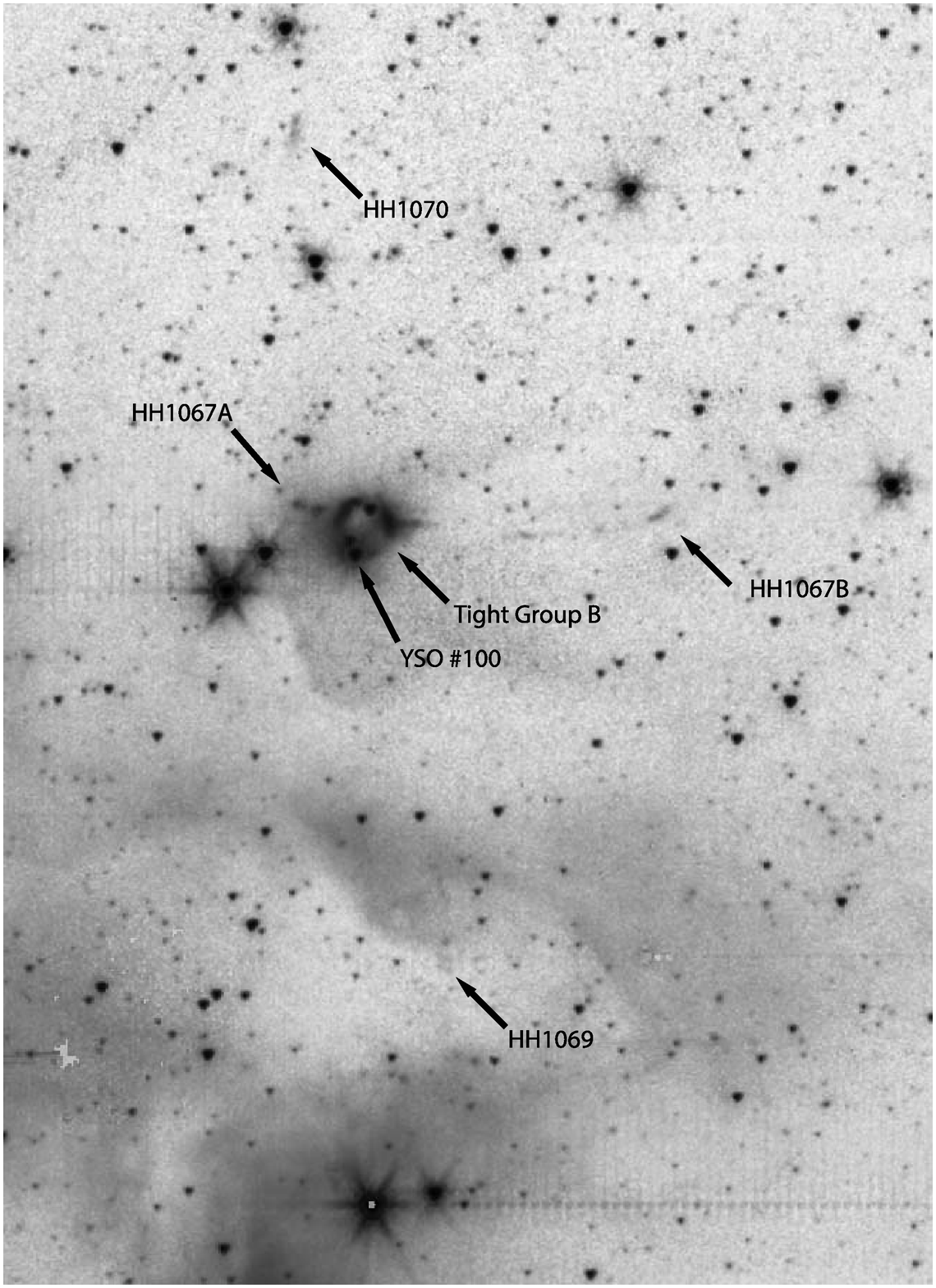}{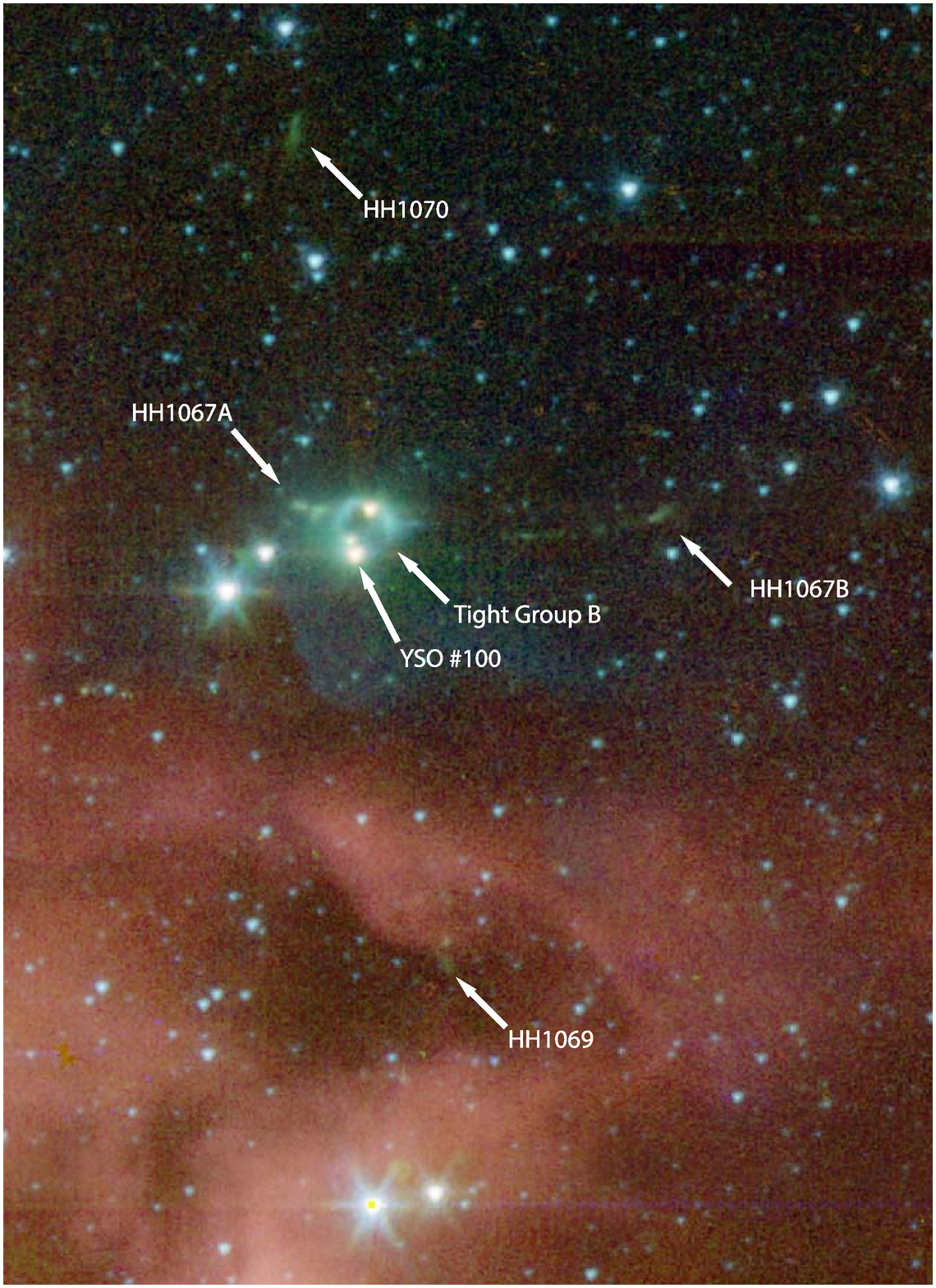}
\caption{A {\it Spitzer} IRAC infrared color composite image of the western portion of NGC~7023 that contains the four newly discovered HH objects.  The image was generated from archival data AOR~25037312.
The field of view is 13\farcm1$\times$18\farcm0. North is up, east is to the left.  The HH objects are labeled, as are the locations of Tight Group B and YSO candidate \#100 from \citet{kir09}.   See the electronic 
edition of the Journal for a color version of this figure.  In the color version, the color assignments for the filters are: 3.6\micron\ (blue), 4.5\micron\ (green), 5.8\micron\ (orange) and 8\micron\ (red).
\label{fig-4}}
\end{figure}

No YSO progenitors are seen in the optical.  However, \citet{kir09} completed a survey of the Cepheus Flare for YSO candidates with the {\it Spitzer} IRAC and MIPS cameras.  In the western lobe of NGC~7023 they identified ``Tight Group B" in the region of our newly discovered outflows.  The group consists of four YSOs  with an associated bubble structure detected at 3.6 and 4.5\micron.  Based upon their infrared spectral indices, \citet{kir09} classify all four of the YSOs in Tight Group B as embedded ``Class I" protostars.  The IRAC archival data (AOR~25037312) of this region are shown in Figure~\ref{fig-4}.  The HH1067 outflow between components A and B, as well as HH1069 and HH1070, are visible at 3.6 and 4.5\micron.  This is consistent with deeply embedded sources, as  emission at these wavelengths is enhanced in molecular shocks at relatively high temperatures or densities as compared to 5.8 and 8\micron\ emission \citep{tak10}.  HH1068 is not in the field of view of the IRAC observations.

HH1067 appears to be a bipolar outflow with an origin in Tight Group B, whose position is marked in Figure~\ref{fig-1}.  In both the optical and infrared, HH1067A and HH1067B are clearly connected by a faintly visible and gently curved segmented line that runs through the group and points towards HH1067C.  HH1067C may be a termination shock, as its structure is nearly perpendicular to the structure of HH1067B and no HH objects are seen to the west of HH1067C.  It is not clear which, if any, of the four YSOs detected in the group is the progenitor for this outflow.    

HH1068, HH1069 and HH1070 are also roughly collinear with Tight Group B.  The structure of HH1070 strongly suggests a bow shock with an origin in the direction of this group.  It is also not clear which YSO is the progenitor of this outflow.  However it is worth noting that HH1069 and HH1070 are equidistant and collinear with YSO \#100 as identified in \citet{kir09}, suggesting that both HH objects were produced by that object as part of the same ejection event.  If HH1068 is also part of this outflow, the angular distance between YSO candidate \#100 and HH1068 is 357\arcsec.  Assuming a distance to NGC~7023 of 430pc, this is a projected distance of 0.75pc, which is a notable length for an embedded outflow.  Assuming a transverse motion of 100 km s$^{-1}$ of the outflow implies a dynamical age of about 7500 yr.  If HH1068, HH1069 and HH1070 are components of a bipolar outflow, then the YSO is likely precessing in a manner similar to \object{PV Cep} \citep{gom97}.  The HH1068-HH1070 outflow appears to be embedded within or behind the reflection nebulosity of NGC~7023, as both HH1069 and HH1070 are visible in cavities.  HH1068 and the western half of HH1070 are fainter and appear to be partially obscured by dust and gas visible in the broadband filters.

The overall structure observed in NGC~7023 is indicative of two distinct outflows with a common origin.  It is similar in appearance to the \object{HH576/577} and \object{BRC 25} cloud core  \citep{rei04}.  Since they are seen as nearly perpendicular, it is hard to envision a scenario where HH1067 is produced by the same progenitor as HH1068-70, although it could be a binary system.  Because HH1067 is visible along most of the outflow, it suggests that it is on the near edge of the NGC~7023 nebula and therefore is likely closer than the HH1068-HH1070 outflow.

\subsection{B175}

\begin{figure}
\plottwo{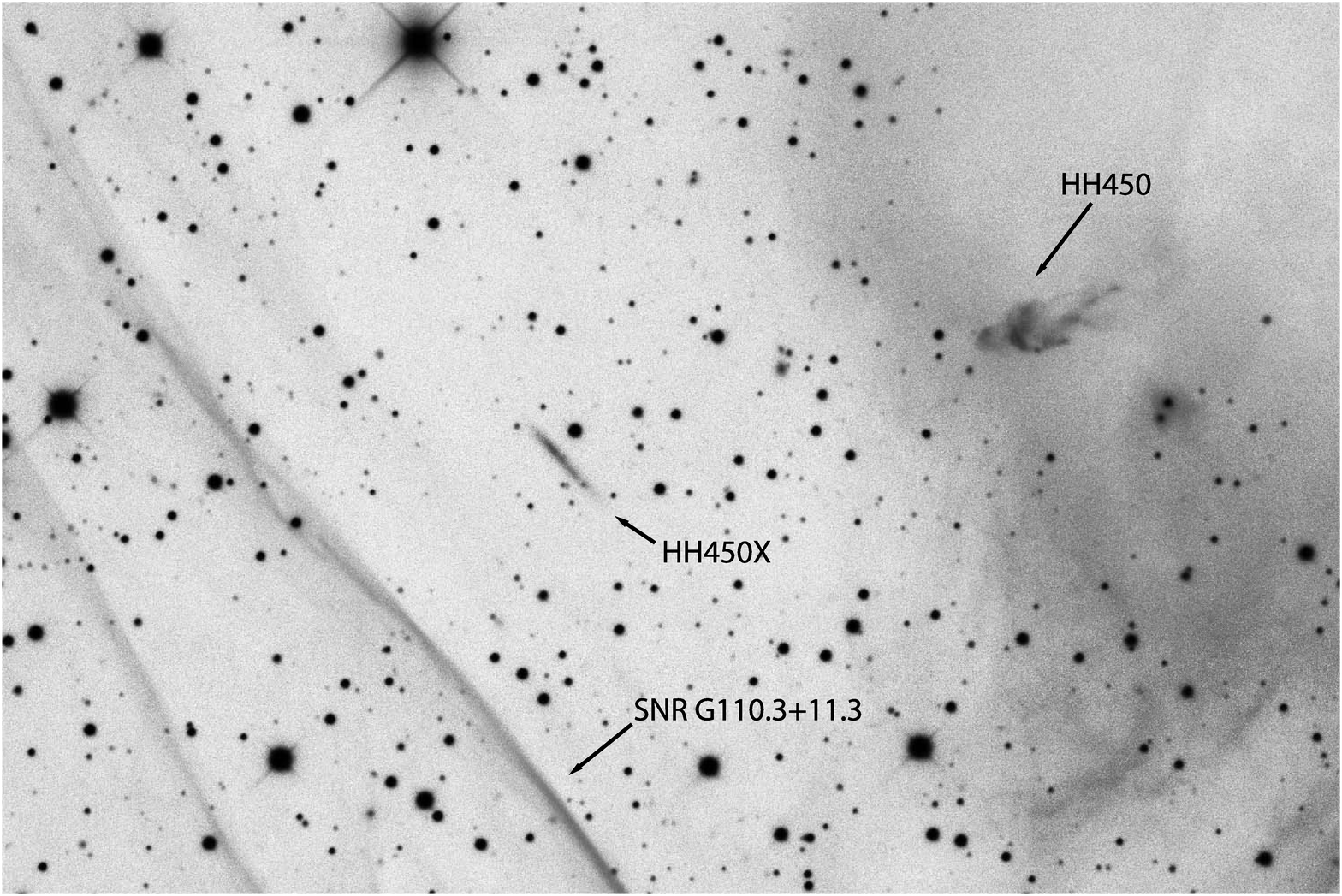}{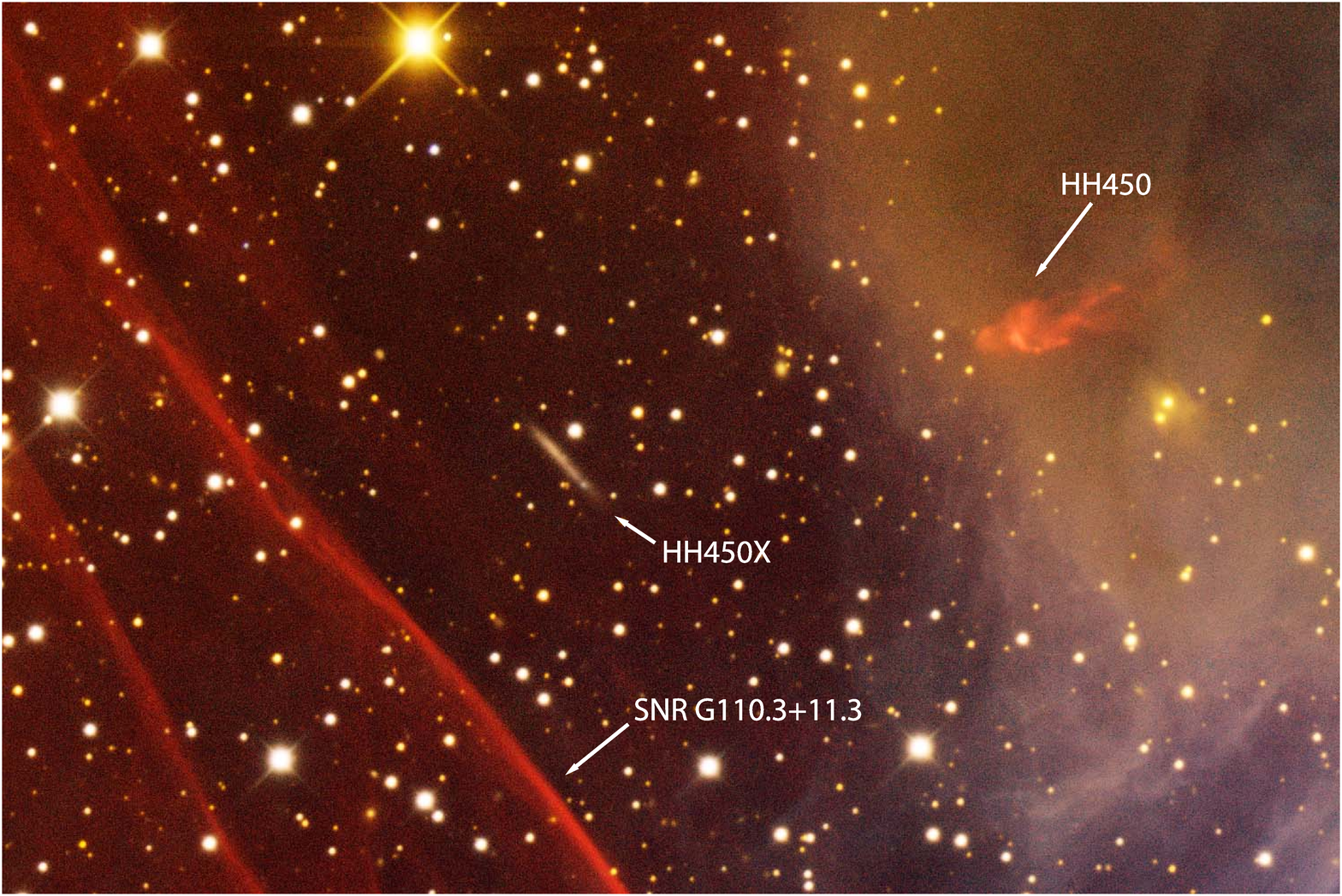}
\caption{A color composite image of the region around HH450X in B175.  The field of view is 4\farcm3$\times$6\farcm5.  North is up, east is to the left.  See the electronic 
edition of the Journal for a color version of this figure.  In the color version, the color assignments for the filters are: $B$ (blue), $V$ (green), $R$ (orange) and H$\alpha$ (red).  HH450X is detected in all filters, and has the color and morphology of a background galaxy.
\label{fig-5}}
\end{figure}

A region 40\farcm6 $\times$ 36\farcm3 in size, centered on the location ${\rm \alpha}_{2000} =22^{\rm h}13^{\rm m}33\fs3, {\rm \delta}_{2000} = +70\degr15\arcmin25\arcsec$ was searched for new HH objects.  The field of B175 is particularly interesting because it contains the outflow \object{HH 450} as well as the supernova remnant \object{SNR G110.3+11.3}, both discovered by \citet{bal01b}.  No new HH objects were found in B175.  However, we do reclassify a small filament located between HH~450 and two large filaments of SNR~G110.3+11.3.  \citet{bal01b} originally classified this filament as HH~450X and interpret it as evidence of interaction between the HH outflow and the supernova remnant.  Unlike HH~450 and SNR~G110.3+11.3, which are detected only in H$\alpha$, we detect HH~450X in all of the broadband and narrowband filters, indicating it is a continuum source.  Based upon its colors and morphology (Figure~\ref{fig-5}) we reclassify HH~450X as a background galaxy.  The discovery that HH~450X is not a shock front eliminates a key piece of evidence that there is a physical connection between B175 and SNR~G110.3+11.3 and makes their relationship less clear. 
%	\section{Conclusions}


%% Included in this acknowledgments section are examples of the
%% AASTeX hypertext markup commands. Use \url without the optional [HREF]
%% argument when you want to print the url directly in the text. Otherwise,
%% use either \url or \anchor, with the HREF as the first argument and the
%% text to be printed in the second.

\acknowledgments

We are grateful to B. Reipurth and J. Bally for their helpful discussions.  We also wish to thank the anonymous referees as well as Kitt Peak National Observatory and its excellent support staff.  The figures in this paper were created with the help of the ESA/ESO/NASA FITS Liberator.


%% in parentheses, after the keyword, but they will not be verified.

{\it Facilities:} \facility{Mayall}.




\clearpage

%% the file, uncomment the \plotone, \plottwo, and
%% If you need a layout that cannot be achieved with \plotone or
%% \plottwo, you can invoke the graphicx package directly with the
%% \includegraphics command or use \plotfiddle. For more information,
%% please see the tutorial on "Using Electronic Art with AASTeX" in the
%% documentation section at the AASTeX Web site,
%% http://www.journals.uchicago.edu/AAS/AASTeX.
%%
%% The examples below also include sample markup for submission of
%% supplemental electronic materials. As always, be sure to check
%% the instructions to authors for the journal you are submitting to
%% for specific submissions guidelines as they vary from

%% This example uses \plotone to include an EPS file scaled to
%% 80% of its natural size with \epsscale. Its caption
%% has been written to indicate that additional figure parts will be
%% available in the electronic journal.

%% Here we use \plottwo to present two versions of the same figure,
%% one in black and white for print the other in RGB color
%% for online presentation. Note that the caption indicates
%% that a color version of the figure will be available online.

%% **NOTE:  Fig 1bw was produced in photoshop via adjustments/black & white, 
%% turning up the red to 300%, inverting the image, and then using curves.

\clearpage

%% Figures 2,3 shows the ha, s2 insets for each HH object.

\clearpage

%% Figure 4 shows the bw, color image of the Spitzer data.

\clearpage
	
%% Figure 5 shows the bw, color image of HH450X and region in B175.

\clearpage

\begin{deluxetable}{lll}
\tablecaption{Identified HH objects in NGC~7023\label{tbl-1}}
\tablewidth{0pt}
\tablehead{\colhead{ID} & \colhead{RA(2000)} & \colhead{DEC}}
\startdata
HH1067A & 21:00:24.5 & +68:13:19 \\
HH1067B & 20:59:59.0 & +68:13:04 \\
HH1067C & 20:59:50.4 & +68:13:21 \\
HH1068A & 21:00:17.6 & +68:18:58 \\
HH1068B & 21:00:14.8 & +68:19:12 \\
HH1069 & 21:00:17.9 & +68:10:13 \\
HH1070 & 21:00:23.7 & +68:16:04 \\
\enddata
\end{deluxetable}

%% The following command ends your manuscript. LaTeX will ignore any text

\end{document}